\documentclass{JAC2003}
\usepackage{graphicx}
\usepackage{booktabs}

\begin{document}
\title{OPTIMAL  TWISS PARAMETERS FOR EMITTANCE\\
MEASUREMENT IN PERIODIC TRANSPORT CHANNELS
\vspace{-0.6cm}}

\author{V.Balandin, W.Decking, 
N.Golubeva\thanks{nina.golubeva@desy.de}\\
DESY, Hamburg, Germany}

\maketitle

\begin{abstract}
\vspace{-0.1cm}
This paper is the continuation of the paper~\cite{PartI},
where we started the study of the impact of the errors
in the beam size measurements on the precision of the
reconstruction of the beam parameters.
Our objective in this paper
is to describe an invariant optimality criterion
and then apply it to the procedure of emittance measurement in 
periodic beam transport channels.
We use, without further explanations, all definitions and 
notations given in~\cite{PartI}, and refer to the 
equations of that paper in the form $(1.*)$.
\end{abstract}

\vspace{-0.2cm}
\section{Optimality Criterion}

\vspace{-0.1cm}
Let us assume that, by using the least squares approach (1.21)-(1.22),
we have obtained the estimate $\mathbf{m}_{\varsigma}(r)$
of the beamlike vector
$\mathbf{m}_0(r)= \varepsilon_0 \,\mathbf{t}_0(r)$ 
actually matched to the
measurements system, and let our estimate be itself
a beamlike vector. Then the estimates $\varepsilon_{\varsigma}$ and 
$\mathbf{t}_{\varsigma}(r)$ for the beam
parameters can be calculated as follows:

\vspace{-0.2cm}
\noindent
\begin{eqnarray}
\varepsilon_{\varsigma}^2 = 
\mathbf{m}_{\varsigma}^{\top}(r)\, S \,\mathbf{m}_{\varsigma}(r),
\quad
\mathbf{t}_{\varsigma}(r) = 
\mathbf{m}_{\varsigma}(r) \,/\, \varepsilon_{\varsigma}. 
\label{Estim02}
\end{eqnarray}

\vspace{-0.2cm}
\noindent
Using the error vector (1.26),
the equation for $\varepsilon_{\varsigma}^2$ can be rewritten in the form

\vspace{-0.2cm}
\noindent
\begin{eqnarray}
\varepsilon_{\varsigma}^2 = \varepsilon_0^2 +
(\mathbf{m}_0^{\top} S \tilde{\mathbf{m}}_{\varsigma} + 
\tilde{\mathbf{m}}_{\varsigma}^{\top} S \mathbf{m}_0) +
\tilde{\mathbf{m}}_{\varsigma}^{\top} S \tilde{\mathbf{m}}_{\varsigma},
\label{SSS01}
\end{eqnarray}

\vspace{-0.2cm}
\noindent
and for the mismatch between
$\mathbf{t}_0$ and $\mathbf{t}_{\varsigma}$ one obtains

\vspace{-0.2cm}
\noindent
\begin{eqnarray}
\varepsilon_{\varsigma} m_p(\mathbf{t}_0, \mathbf{t}_{\varsigma}) =
\varepsilon_0
+
\left(\mathbf{m}_0^{\top} S \tilde{\mathbf{m}}_{\varsigma} + 
\tilde{\mathbf{m}}_{\varsigma}^{\top} S \mathbf{m}_0\right)
/ (2 \varepsilon_0).
\label{SSS02}
\end{eqnarray}

\vspace{-0.2cm}
\noindent
Averaging of both sides of the formulas (\ref{SSS01})
and (\ref{SSS02}) with respect to the measurement statistics
gives us 

\vspace{-0.2cm}
\noindent
\begin{eqnarray}
\langle
\varepsilon_{\varsigma}^2 
\rangle 
= \varepsilon_0^2 +
\langle
\tilde{\mathbf{m}}_{\varsigma}^{\top} S \tilde{\mathbf{m}}_{\varsigma}
\rangle 
= \varepsilon_0^2 +
\mbox{tr}(V_m S),
\label{SSS03}
\end{eqnarray}

\vspace{-0.4cm}
\noindent
\begin{eqnarray}
\langle
\varepsilon_{\varsigma}
m_p(\mathbf{t}_0, \mathbf{t}_{\varsigma})
\rangle 
= \varepsilon_0.
\label{SSS04}
\end{eqnarray}

\vspace{-0.2cm}
\noindent
These formulas are quite remarkable.
They are exact, they involve only first (which are equal to zero in our case) 
and second statistical moments of 
the error vector $\tilde{\mathbf{m}}_{\varsigma}$, 
and they do not depend on the matched Twiss vector $\mathbf{t}_0$.
Besides that, one sees the appearance of the invariant of the 
covariance matrix $V_m$ in the right hand side of the equation (\ref{SSS03}).
Due to (1.34) this invariant is negative if the matrix $V_{\varsigma}$
is diagonal. It means that if errors in the beam size measurements are
uncorrelated, then the 
``typical reconstructed emittance" $\varepsilon_{\varsigma}$
is an underestimation of the real emittance $\varepsilon_0$,
that was already observed a number of times  
by different authors in their Monte-Carlo simulations.

To go further, behind the exact relations (\ref{SSS03}) and (\ref{SSS04}), 
we do need to make some approximation. Let us assume that the
measurement errors are not too large and
let us denote

\vspace{-0.2cm}
\noindent
\begin{eqnarray}
{\cal A} = 
\frac{\mathbf{m}_0^{\top} S \,\tilde{\mathbf{m}}_{\varsigma} + 
\tilde{\mathbf{m}}_{\varsigma}^{\top} S \,\mathbf{m}_0}{2 \varepsilon_0^2},
\quad
{\cal B} = 
\frac{\tilde{\mathbf{m}}_{\varsigma}^{\top} S\, \tilde{\mathbf{m}}_{\varsigma}}
{\varepsilon_0^2}.
\label{NV01}
\end{eqnarray}

\vspace{-0.2cm}
\noindent
With these notations we obtain from (\ref{SSS01}) and (\ref{SSS02})
that

\vspace{-0.2cm}
\noindent
\begin{eqnarray}
\frac{\varepsilon_{\varsigma} - \varepsilon_0}{\varepsilon_0}
= 
\sqrt{(1 + {\cal A})^2 - ({\cal A}^2 - {\cal B})} - 1,
\label{EstEmit2}
\end{eqnarray}

\vspace{-0.4cm}
\noindent
\begin{eqnarray}
d_H(\mathbf{t}_{\varsigma}, \mathbf{t}_0)
= \mbox{arccosh}
\left(
\frac{1 + {\cal A}} 
{\sqrt{(1 + {\cal A})^2 - ({\cal A}^2 - {\cal B})}}
\right).
\label{EstMism}
\end{eqnarray}

\vspace{-0.2cm}
\noindent
Squaring both sides of these formulas and then
expanding the right hand sides 
of the obtained equalities
with respect to the variables ${\cal A}$ and
${\cal A}^2 - {\cal B}$, we obtain 

\vspace{-0.2cm}
\noindent
\begin{eqnarray}
\left(
(\varepsilon_{\varsigma} - \varepsilon_0) \,/\,\varepsilon_0  
\right)^2
=
{\cal A}^2 + \ldots,
\label{EstEmit22}
\end{eqnarray}

\vspace{-0.4cm}
\noindent
\begin{eqnarray}
d_H^2(\mathbf{t}_{\varsigma}, \mathbf{t}_0)
=
{\cal A}^2 - {\cal B} + \ldots,
\label{EstEmit223}
\end{eqnarray}

\vspace{-0.2cm}
\noindent
where the omitted terms are of the orders three and higher
in the components of the error vector
$\tilde{\mathbf{m}}_{\varsigma}$.

We take the functions

\vspace{-0.2cm}
\noindent
\begin{eqnarray}
\psi_1
=
{\cal A}^2
\quad
\mbox{and}
\quad
\psi_2
=
{\cal A}^2 - {\cal B}
\label{EstEmit2237}
\end{eqnarray}

\vspace{-0.2cm}
\noindent
as the basic components for the construction
of the vector-valued optimality criterion.
Both these functions can be written as
quadratic forms with respect to the vector
$\tilde{\mathbf{m}}_{\varsigma}$:

\vspace{-0.2cm}
\noindent
\begin{eqnarray}
\psi_1
=
(1 \,/\, {\varepsilon_0^2}) \cdot
\tilde{\mathbf{m}}_{\varsigma}^{\top}
\left(
S \, \mathbf{t}_0 \mathbf{t}_0^{\top} S 
\right)
\tilde{\mathbf{m}}_{\varsigma},
\label{QuadForm01}
\end{eqnarray}

\vspace{-0.4cm}
\noindent
\begin{eqnarray}
\psi_2
=
(1 \,/\, {\varepsilon_0^2}) \cdot
\tilde{\mathbf{m}}_{\varsigma}^{\top}
\left(
S \, \mathbf{t}_0 \mathbf{t}_0^{\top} S - S
\right)
\tilde{\mathbf{m}}_{\varsigma}.
\label{QuadForm02}
\end{eqnarray}

\vspace{-0.2cm}
\noindent
Using this representation, one finds that
each of these quadratic forms is positive semidefinite,
but is not positive definite. Moreover, it is possible 
to show that $\psi_1$ and $\psi_2$ are incomparable
(there are points where $\psi_1$ is equal to zero, 
but $\psi_2$ is not, and vice versa),
i.e., really, properties of both these functions have to be reflected
in the optimality criterion.\footnote{If only measurement of emittance 
is of interest, then, looking at the equation (\ref{EstEmit22}),
one may conclude that the optimization of $\psi_1$ alone 
could be sufficient. In general, it is not true. The function
$\psi_2$, if not controlled, may become very large
and can spoil the optimization result through the high order terms 
which are not shown in (\ref{EstEmit22}).}
The values of the functions $\psi_1$ and $\psi_2$
do not depend on the positioning of the reconstruction point,
and as the components of the optimality criterion we take
the averages

\vspace{-0.2cm}
\noindent
\begin{eqnarray}
\langle
\psi_1
\rangle =  
{\cal F} \,/\,\varepsilon_0^2,
\quad
\langle
\psi_2
\rangle =  
(
{\cal F}
- \mbox{tr}(V_m S))
\,/\,\varepsilon_0^2,
\label{AV2}
\end{eqnarray}

\vspace{-0.2cm}
\noindent
where the function ${\cal F}$ is defined in (1.35).

If the errors in the beam size determination
at different measurement states
are uncorrelated, then one can write 

\vspace{-0.2cm}
\noindent
\begin{eqnarray}
\langle
\psi_1
\rangle
=
\frac{1}{2 \varepsilon_0^2 \Delta_{\varsigma}} \sum_{i,j = 1}^{n} 
\frac{a_{12}^2(s_i, s_j)}
{\sigma_i \,\sigma_j} \cdot
\frac{\beta_0(s_i)}{\sigma_i} \cdot
\frac{\beta_0(s_j)}{\sigma_j}
\nonumber
\end{eqnarray}

\vspace{-0.4cm}
\noindent
\begin{eqnarray}
-\frac{1}{2 \varepsilon_0^2 \Delta_{\varsigma}} \sum_{i,j = 1}^{n} 
\left(\frac{a_{12}^2(s_i, s_j)}
{\sigma_i \,\sigma_j}
\right)^2 
= 
\nonumber
\end{eqnarray}

\vspace{-0.4cm}
\noindent
\begin{eqnarray}
\frac{1}{8 \varepsilon_0^2 \Delta_{\varsigma}}
\sum_{i,j = 1}^{n}
\left(\sin(2 \mu_0(s_i, s_j)) \cdot 
\frac{\beta_0(s_i)}{\sigma_i} \cdot
\frac{\beta_0(s_j)}{\sigma_j}\right)^2,
\label{mM3786}
\end{eqnarray}

\vspace{-0.4cm}
\noindent
\begin{eqnarray}
\langle
\psi_2
\rangle
=
\frac{1}{2 \varepsilon_0^2 \Delta_{\varsigma}} \sum_{i,j = 1}^{n} 
\frac{a_{12}^2(s_i, s_j)}
{\sigma_i \,\sigma_j} \cdot
\frac{\beta_0(s_i)}{\sigma_i} \cdot
\frac{\beta_0(s_j)}{\sigma_j}
=
\nonumber
\end{eqnarray}

\vspace{-0.4cm}
\noindent
\begin{eqnarray}
\frac{1}{2 \varepsilon_0^2 \Delta_{\varsigma}}
\sum_{i,j = 1}^{n}
\left(\sin(\mu_0(s_i, s_j)) \cdot 
\frac{\beta_0(s_i)}{\sigma_i} \cdot
\frac{\beta_0(s_j)}{\sigma_j}\right)^2,
\label{mM3787}
\end{eqnarray}

\vspace{-0.2cm}
\noindent
where $\beta_0$ is the betatron
function matched to the measurement system and
$\mu_0$ is the corresponding phase advance.

The other useful forms of $\langle \psi_1 \rangle$
and $\langle \psi_2 \rangle$ can be obtained if we
will use the notations

\vspace{-0.1cm}
\noindent
\begin{eqnarray}
\begin{array}{c}
{\cal C}_m(r)\\
{\cal S}_m(r)
\end{array}
=
\sum_{k = 1}^{ n } 
\left(\frac{\beta_0(s_k)}{\sigma_k}\right)^2
\cdot
\begin{array}{c}
\cos\left(m \mu_0(r, s_k)\right)\\
\sin\left(m \mu_0(r, s_k)\right)
\end{array}
\label{SEC26}
\end{eqnarray}

\vspace{-0.2cm}
\noindent
With these notations:

\vspace{-0.2cm}
\noindent
\begin{eqnarray}
\langle
\psi_1
\rangle
=
\left( 
{\cal C}_0^2 - {\cal C}_4^2 - {\cal S}_4^2
\right) \,/\,
\left(
16 \, \varepsilon_0^2 \,\Delta_{\varsigma}
\right),
\label{SEC219}
\end{eqnarray}

\vspace{-0.4cm}
\noindent
\begin{eqnarray}
\langle
\psi_2
\rangle
=
\left( 
{\cal C}_0^2 - {\cal C}_2^2 - {\cal S}_2^2
\right) \,/\,
\left(
4 \, \varepsilon_0^2 \,\Delta_{\varsigma}
\right),
\label{SEC291}
\end{eqnarray}

\vspace{-0.2cm}
\noindent
where $\Delta_{\varsigma}$ can now be expressed as follows

\vspace{-0.2cm}
\noindent
\begin{eqnarray}
16 \, \Delta_{\varsigma} =
{\cal C}_0 \cdot
\left[
{\cal C}_0^2
-  {\cal C}_4^2 - {\cal S}_4^2
\right]
\nonumber
\end{eqnarray}

\vspace{-0.4cm}
\noindent
\begin{eqnarray}
- 2 
\left[
({\cal C}_0 - {\cal C}_4)\cdot{\cal C}_2^2 
-2\, {\cal S}_4\cdot {\cal C}_2\, {\cal S}_2
+ ({\cal C}_0 + {\cal C}_4)\cdot{\cal S}_2^2 
\right].
\label{SEC217}
\end{eqnarray}

\vspace{-0.2cm}
Let us introduce vector
$\tilde{\varsigma} = V_{\varsigma}^{-1/2} \varsigma$,
where $\varsigma$
is the vector of the measurement errors.
The vector $\tilde{\varsigma}$ has 
covariance matrix equal to the identity matrix, 
that allows us to say that its components are
``better balanced in the order of magnitude"
than the components of the original vector of the measurement errors.
Using the vector $\tilde{\varsigma}$, the expression for the functions $\psi_m$
($m=1,2$) can be written in the form

\vspace{-0.1cm}
\noindent
\begin{eqnarray}
{\varepsilon_0^2} \,\psi_m 
=
\tilde{{\varsigma}}^{\top}
\big(
V_{\varsigma}^{-1/2}
\big[
W - (m-1)\, U
\big]
V_{\varsigma}^{-1/2}
\big)\,
\tilde{\varsigma},
\label{QuadForm03}
\end{eqnarray}

\vspace{-0.1cm}
\noindent
where the matrices $U$ and $W$
are defined in the equations (1.37) and (1.38), respectively.
One can calculate, that

\vspace{-0.1cm}
\noindent
\begin{eqnarray}
\mbox{tr}
\big(
V_{\varsigma}^{-1/2}
\big[
W - (m-1)\, U
\big]
V_{\varsigma}^{-1/2}
\big)
=
{\varepsilon_0^2} \,
\langle 
\psi_m 
\rangle,
\label{QuadForm04}
\end{eqnarray}

\vspace{-0.1cm}
\noindent
and, therefore, one can obtain the estimates

\vspace{-0.1cm}
\noindent
\begin{eqnarray}
0 \,\leq\, \psi_m
\,\leq\,
\langle 
\psi_m 
\rangle
\cdot
\tilde{{\varsigma}}^{\top}
\tilde{{\varsigma}},
\quad
m = 1,2.
\label{QuadForm06}
\end{eqnarray}

\vspace{-0.2cm}
\noindent
The upper estimates of this type are usually called
worst-case estimates, and we see that the minimization of
the statistical averages of the functions $\psi_m$
improve also their upper worst-case 
estimates.\footnote{The more precise upper estimate of a
positive semidefinite 
quadratic form
can be obtained if one will use not the trace (sum of eigenvalues),
but the largest eigenvalue of its matrix.
Unfortunately, it is not easy to find
these largest eigenvalues analytically for the quadratic forms
(\ref{QuadForm03}), while the traces can easily be calculated.}

So we have introduced the optimality criterion, which is independent
from the position of the reconstruction point.
Unfortunately, this criterion includes two objective functions.
To reduce a number of
objectives to a single one 
we suggest to use the additive convolution and take the average of 
the weighted sum 

\vspace{-0.2cm}
\noindent
\begin{eqnarray}
\psi_0 = \kappa_1 \,\psi_1 + \kappa_2 \,\psi_2, 
\quad
\kappa_{1,2} > 0,
\label{QuadForm070}
\end{eqnarray}

\vspace{-0.2cm}
\noindent
as the single valued optimality criterion:

\vspace{-0.2cm}
\noindent
\begin{eqnarray}
\langle 
\psi_0 
\rangle
\,=\,
(
(\kappa_1 + \kappa_2)\,{\cal F}
- \kappa_2\,\mbox{tr}(V_m S))
\,/\,\varepsilon_0^2.
\label{QuadForm07}
\end{eqnarray}

\vspace{-0.2cm}
Note that, for arbitrary positive $\kappa_1$ and $\kappa_2$,
the function $\psi_0$ is a positive definite quadratic form
with respect to the error vector 
$\tilde{\mathbf{m}}_{\varsigma}$. It 
is also a leading term of the expansion of the function

\vspace{-0.2cm}
\noindent
\begin{eqnarray}
\Psi_0 = 
\kappa_1 \, (\varepsilon_{\varsigma} - \varepsilon_0)^2 /\,
(\varepsilon_{\varsigma}\, \varepsilon_0)  
+ \kappa_2 \, d_H^2(\mathbf{t}_{\varsigma}, \mathbf{t}_0)
\label{QFINTR01}
\end{eqnarray}

\vspace{-0.2cm}
\noindent
with respect to the components of the same vector
$\tilde{\mathbf{m}}_{\varsigma}$. 
Thus, as an output of Monte-Carlo simulations,
one may use the average $\langle \Psi_0 \rangle$
and compare it with the analytical predictions 
given by the function $\langle \psi_0 \rangle$.

The choice of the weights $\kappa_1$ and $\kappa_2$
(as usual in the area of the multicriteria analysis)
should reflect the specific of the problem
under study, but, as concerning general situation,
we do not see currently any clear theoretical reasons to take them unequal.

\vspace{-0.01cm}
\section{Optimal Twiss Parameters}

\vspace{-0.01cm}
The Twiss parameters matched to the measurement system 
enter the optimality criterion in two
different ways, directly through the function ${\cal F}$
and indirectly, when the measurement errors
depend on the measured beam sizes (i.e when the matrix
$V_{\varsigma}$ depend on the vector $\mathbf{b}_0$). 

If the matrix $V_{\varsigma}$ depend on the vector $\mathbf{b}_0$,
then there is not much that one can say besides the trivial statement
that the optimal Twiss parameters are the Twiss parameter which,
for the fixed transport matrices between measurement states,
minimize the optimality criterion.

So, let us assume that the measurement errors
are independent from the measured beam sizes.
In this situation one can prove that

\vspace{-0.2cm}
\noindent
\begin{eqnarray}
\min_{\mathbf{t}_0 \in {\cal T}_s}\, {\cal F} \;=\;
1 \,/\, \min_{\mathbf{t}_0 \in {\cal T}_s}\, {\cal G} 
\;=\;
\lambda_1,
\label{INVQF03}
\end{eqnarray}

\vspace{-0.2cm}
\noindent
where ${\cal G}$ is defined in (1.36).
From this it follows that, if the Twiss parameters are optimal, then 

\vspace{-0.2cm}
\noindent
\begin{eqnarray}
\langle \psi_1 \rangle = \lambda_1 \,/\, \varepsilon_0^2
\quad
\mbox{and}
\quad
\langle \psi_2 \rangle
= |\lambda_2 + \lambda_3|\,/\, \varepsilon_0^2.
\label{OPTLAM}
\end{eqnarray}

\vspace{-0.2cm}
Let us assume additionally that the matrix
$V_{\varsigma}$ is diagonal.
Then the second equality in (\ref{INVQF03}) takes on the form

\vspace{-0.2cm}
\noindent
\begin{eqnarray}
\frac{1}{\lambda_1} =
\min_{\mathbf{t}_0 \in {\cal T}_s}\,
\sum_{i = 1}^{n}
\left( 
\frac{\beta_0(s_i)}{\sigma_i}
\right)^2,
\label{INVQF05}
\end{eqnarray}

\vspace{-0.2cm}
\noindent
i.e. the optimal Twiss parameters minimize 
the weighted sum of squares of the betatron function.

Because both minimums in (\ref{INVQF03}) are achieved 
in the same point,
the optimal Twiss parameters (and only they) satisfy

\vspace{-0.2cm}
\noindent
\begin{eqnarray}
{\cal F} = {\cal C}_0^{-1}.
\label{SEC2_31}
\end{eqnarray}

\vspace{-0.2cm}
\noindent
Using the formulas (\ref{SEC219}) and (\ref{SEC217}),
the equality (\ref{SEC2_31}) can be transformed into the following
equivalent form

\vspace{-0.2cm}
\noindent
\begin{eqnarray}
({\cal C}_0 - {\cal C}_4) \cdot {\cal C}_2^2
-2 {\cal S}_4 \cdot {\cal C}_2 {\cal S}_2 
+({\cal C}_0 + {\cal C}_4) \cdot {\cal S}_2^2 = 0.
\label{SEC2_32}
\end{eqnarray}

\vspace{-0.2cm}
\noindent
The left hand side of (\ref{SEC2_32})
is a positive definite quadratic form in the
variables ${\cal C}_2$ and ${\cal S}_2$, and therefore
it can be equal to zero if and only if

\vspace{-0.2cm}
\noindent
\begin{eqnarray}
{\cal C}_2 = {\cal S}_2 = 0,
\label{SEC2_35}
\end{eqnarray}

\vspace{-0.2cm}
\noindent
which is the characteristic property of the optimal Twiss parameters.


\begin{figure}[!htb]
    \centering
    \includegraphics*[width=65mm]{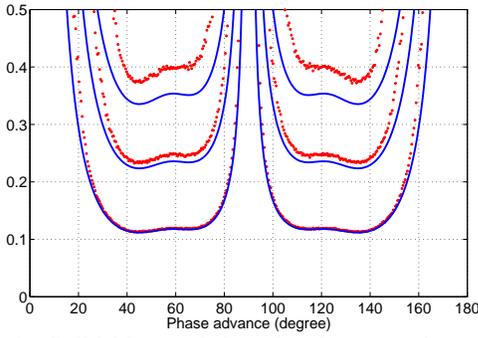}
    \vspace{-0.5cm}
    \caption{Solid blue and dotted red curves show
    $\sqrt{\langle \psi_0 \rangle}$ and 
    $\sqrt{\langle \Psi_0 \rangle}$, respectively, 
    as functions of the cell phase advance.
    The measurement errors are relative and are equal to
    $10\%$, $20\%$ and $30\%$ (from lower to upper pairs of curves).} 
    \label{fig1}
    \vspace{-0.3cm}
\end{figure}

\section{Emittance Measurement\\
in Periodic Transport Channels}

\vspace{-0.1cm}
In this section we consider 
the question of optimal phase advance choice  
for the procedure of
emittance measurement in periodic beam transport
channels. We assume that we have a measurement system 
with $n$ measurement states $s_1, \ldots, s_n$ and
with the property that the matrices propagating 
particle coordinates from the state $s_m$ to the
state $s_{m+1}$ are all equal to each other
and allow periodic beam transport with the periodic 
betatron function $\beta_p$ and with the corresponding phase advance
$\mu_p$ being not a multiple of $90^{\circ}$. 
We assume also that this periodic betatron function
is the betatron function matched to the measurement system
and that the errors in the beam size
determination at different measurement states are uncorrelated
and have the same rms magnitude $\sigma_p$ 
(i.e. $V_{\varsigma} = \sigma_p^2 \, I$).

Under these assumptions the formulas (\ref{SEC219})-(\ref{SEC291}) 
for the components of the optimality criterion can be rewritten as follows:

\vspace{-0.3cm}
\noindent
\begin{eqnarray}
\langle \psi_1 \rangle =
\frac{1}{n}
\left(
\frac{\sigma_p}{\varepsilon_0 \, \beta_p}
\right)^2
\cdot \varrho_n(\mu_p),
\label{PMS01}
\end{eqnarray}

\vspace{-0.4cm}
\noindent
\begin{eqnarray}
\langle \psi_2 \rangle =
\frac{4}{n}
\left(
\frac{\sigma_p}{\varepsilon_0 \, \beta_p}
\right)^2
\cdot \varpi_n(\mu_p),
\label{PMS03}
\end{eqnarray}

\vspace{-0.25cm}
\noindent
where

\vspace{-0.25cm}
\noindent
\begin{eqnarray}
\varrho_n(\mu_p) =
\frac{1 + \mbox{\ae}_n (2 \mu_p)}
{1 + \mbox{\ae}_n (2 \mu_p) - 2 \,\mbox{\ae}_n^2 (\mu_p)},
\label{PMS02}
\end{eqnarray}

\vspace{-0.4cm}
\noindent
\begin{eqnarray}
\varpi_n(\mu_p) =
\frac{1 - \mbox{\ae}_n^2 (\mu_p)}
{1 - \mbox{\ae}_n^2 (2 \mu_p)} \cdot
\varrho_n(\mu_p),
\label{PMS04}
\end{eqnarray}

\vspace{-0.25cm}
\noindent
and $\mbox{\ae}_n$ is given by

\vspace{-0.25cm}
\noindent
\begin{eqnarray}
\mbox{\ae}_n(\mu_p) =
\frac{\sin(n \mu_p)}
{n\,\sin(\mu_p)}.
\label{PMS05}
\end{eqnarray}

\vspace{-0.25cm}
\noindent
For an arbitrary $n \geq 3$  the function $\varrho_n(\mu_p)$
is $180^{\circ}$-periodic and
can be extended by continuity for all $\mu_p$ inside a period, and
becomes unbounded as one approaches $0^{\circ}$ and $180^{\circ}$.
It is never smaller than one and is equal
to one (reaches its minimum) only in the points 

\vspace{-0.25cm}
\noindent
\begin{eqnarray}
\mu_p = k \cdot 180^{\circ} \,/\, n \;\;\;(\mbox{mod} \;180^{\circ}),
\label{PMS06}
\end{eqnarray}

\vspace{-0.25cm}
\noindent
where $k = 1, ... , n-1$ with the exception of the value $n / 2$

\newpage

\noindent
if $n$ is even.
The same properties hold also for the function $\varpi_n(\mu_p)$
with the addition that this function 
becomes unbounded also near $90^{\circ}$.

Looking at the formulas (\ref{PMS01}) and (\ref{PMS03}) 
one sees that the choice of $\mu_p$ in accordance with the rule
(\ref{PMS06}) brings the second multipliers in the right hand
sides of these formulas to the minimal possible values.
But, in general, it does not guarantee that the products
of the two multipliers in (\ref{PMS01}) and (\ref{PMS03})
are also minimized. 
So the answer to the question,
if a $n$-cell measurement system reaches an optimal performance when its
cell phase advance is a multiple of $180^{\circ}$ divided by $n$,
depends on the behavior of the ratio $\sigma_p \,/\, (\varepsilon_0 \,\beta_p)$
during the scan of the phase advance. 
If this ratio stays unchanged, then it is certainly true.
It includes, for example, the case when 
$\sigma_p$ is proportional to $\varepsilon_0 \,\beta_p$, i.e. when 
errors in the beam size determination are relative to the beam size itself.
But if the ratio  $\sigma_p \,/\, (\varepsilon_0 \,\beta_p)$ changes, 
then the optimality of the phase advances (\ref{PMS06})
is not guaranteed.

To be more specific, let us consider a transport channel constructed 
from four identical thin lens FODO cells of the unit length and let us 
assume that four measurement stations are placed in the middle of drift 
spaces separating the defocusing and focusing lenses.
For this particular system we made a series of Monte-Carlo simulations
using as an output the square root of the average of the functions 
(\ref{QFINTR01}) with $\kappa_1 = \kappa_2 = 1$, and then
compared them with the analytical predictions given
by the square root of the function (\ref{QuadForm07}). 
These simulations and comparisons were made for two cases:
measurement errors are relative to the beam size
and measurement errors are beam size independent.
The results are presented in Fig.1 and Fig.2, respectively.
One sees, that while for the relative errors the optimal values
of $\mu_p$ are $45^{\circ}$ and $135^{\circ}$
(as expected), the effect of the absolute errors gives
a single optimal point $\mu_p \approx 34.5^{\circ}$.

\vspace{-0.2cm}

\begin{figure}[!t]
    \centering
    \includegraphics*[width=65mm]{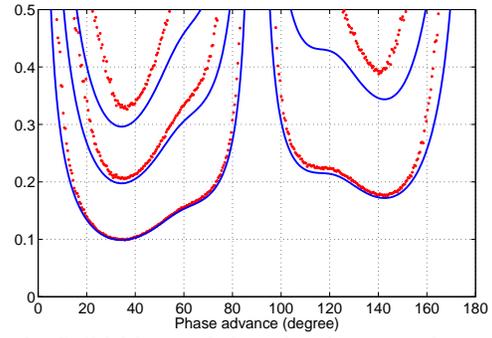}
    \vspace{-0.5cm}
    \caption{Solid blue and dotted red curves show
    $\sqrt{\langle \psi_0 \rangle}$ and 
    $\sqrt{\langle \Psi_0 \rangle}$, respectively, 
    as functions of the cell phase advance.
    The measurement errors are beam size independent and are normalized
    to coincide with the relative errors of
    $10\%$, $20\%$ and $30\%$ for $\mu_p = 45^{\circ}$ (from lower to upper pairs of curves).}  
    \label{fig2}
    \vspace{-0.5cm}
\end{figure}

\end{document}